\documentclass[doublecol]{epl2} 
\usepackage{amsmath}
\usepackage{amssymb}
\usepackage{amsfonts}
\usepackage{color}
\usepackage[utf8]{inputenc}
\usepackage[T1]{fontenc}

\usepackage[usenames,dvipsnames]{xcolor}
\usepackage{amsbsy}
\usepackage{ulem}

\newcommand{\ind}[1]{_\textrm{#1}}
\newcommand{\fext}{K\ind{ext}}
\newcommand{\floc}{{K\ind{c}}}
\newcommand{\Floc}{k\ind{c}}
\newcommand{\fc}{{K\ind{eff}}}

\newcommand{\tfc}{\tilde K\ind{eff}}
\newcommand{\tFc}{\tilde k\ind{eff}}
\newcommand{\Lc}{L\ind{c}}

\definecolor{red}{rgb}{1,0,0}

\newcommand{\dKcloc}{\sigma}

\newcommand{\dd}{\mathrm{d}}

\title{From microstructural features to effective toughness in disordered brittle solids}

\author{V. D\' emery\inst{1}\thanks{E-mail: \email{vincent.demery@polytechnique.edu}} \and A. Rosso\inst{2} \and L. Ponson\inst{1}}
\shortauthor{V. D\' emery \etal}

\institute{                    
  \inst{1} Institut Jean Le Rond d'Alembert (UMR 7190), CNRS and Universit\'e Pierre et Marie Curie, 75005 Paris, France\\
  \inst{2} Laboratoire Physique Th{\'e}orique et Mod{\`e}les  Statistiques (UMR 8626), Universit\'e de Paris-Sud, Orsay Cedex, France}
  

\abstract{The relevant parameters at the microstructure scale that govern the macroscopic toughness of disordered brittle materials are investigated theoretically. We focus on planar crack propagation and describe the front evolution as the propagation of a long-range elastic line within a plane with random distribution of toughness. Our study reveals two regimes: in the collective pinning regime, the macroscopic toughness can be expressed as a function of a few parameters only, namely the average and the standard deviation of the local toughness distribution and the correlation lengths of the heterogeneous toughness field; in the individual pinning regime, the passage from micro to macroscale is more subtle and the full distribution of local toughness is required to be predictive. Beyond the failure of brittle solids, our findings illustrate the complex filtering process of microscale quantities towards the larger scales into play in a broad range of systems governed by the propagation of an elastic interface in a disordered medium.}

\begin{document}

\maketitle

Bridging microscale features of materials with their effective behavior at the macroscale is a major challenge for properties governed by the propagation of interfaces or free boundaries. In systems like brittle solids~\cite{Bonamy6}, ferromagnets~\cite{Zapperi2}, superconductors ~\cite{Larkin1979}, martensitic solids~\cite{Vives}, thin film adhesives~\cite{Xia}, or wetting films~\cite{Moulinet2004}, impurities or defects present at the microstructure scale can produce dramatic macroscopic effects. But they also sometimes have interesting benefits. For example, large precipitate particles trap dislocations in metallic alloys, increasing notably their overall strength~\cite{Callister}.

\begin{figure}
\begin{center}
\includegraphics[width=.9\linewidth]{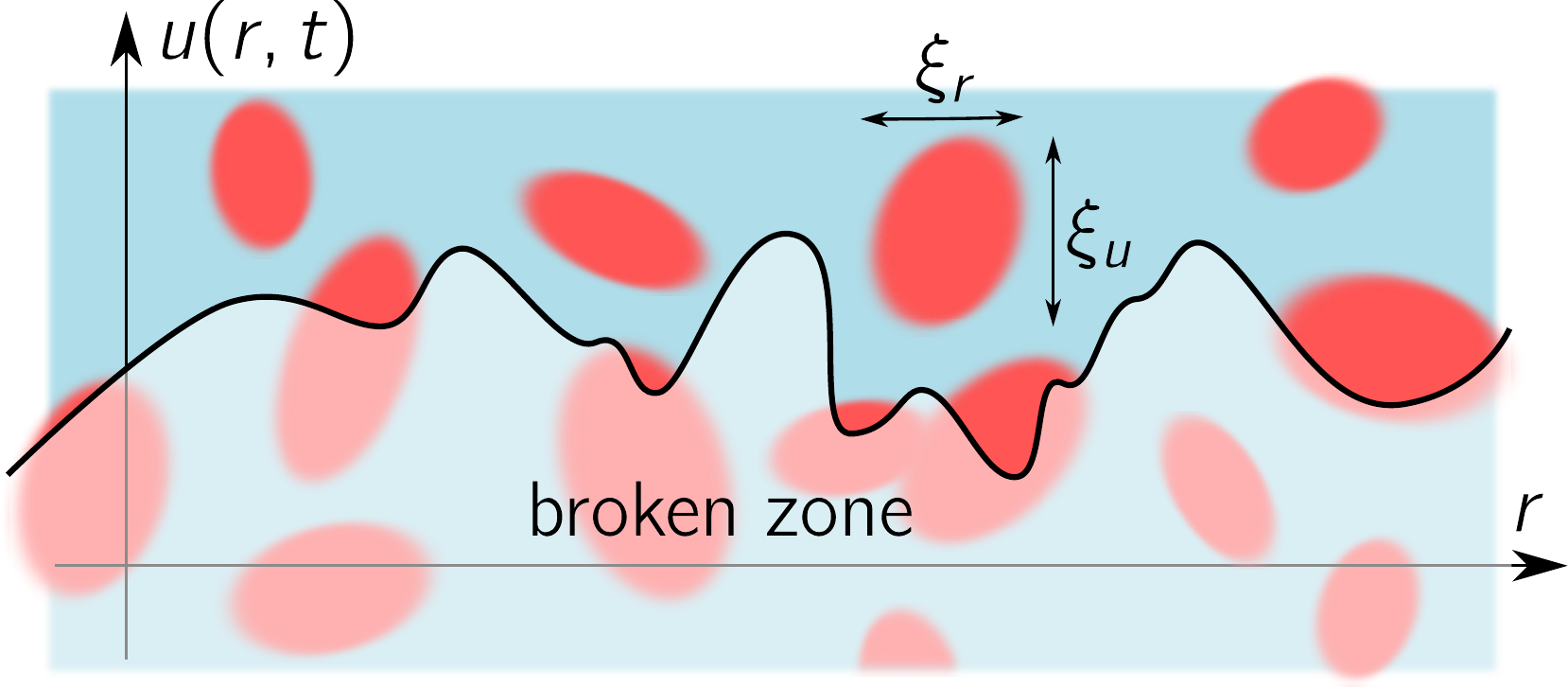}
\caption{(Color online) Crack front moving upwards in a disordered medium: dark red stands for tough areas, and light blue for low toughness areas. The defects have characteristic sizes $\xi_r$ and $\xi_u$ in the $r$ and $u$ directions, respectively.}
\label{schema_gen}
\end{center}
\end{figure}

In this study, we address the challenge of determining the effective toughness of brittle solids from the variations of toughness at their microstructure scale. Failure processes involve in general complex mechanisms such as damage and plastic deformations resulting from the high level of tensile stress in the crack tip vicinity. These mechanisms are localized in the so-called process zone, and a major simplification occurs in brittle solids where the process zone size is much smaller than the typical size of heterogeneities in the material. Assuming a planar crack propagation, the crack propagation can then be modeled as the overdamped motion of a line $u(r,t)$ moving in a two dimensional heterogeneous medium~\cite{Gao,Schmittbuhl4,Ponson17}, as represented on Fig.~\ref{schema_gen}. Previous works have investigated the effect of tough inclusions on the propagation of brittle cracks and were able to quantify the toughening induced by a periodic distribution of these obstacles~\cite{Bower,Gao}. Here, we focus on disordered microstructures, like randomly distributed impurities, from which emerge new toughening mechanisms.

Earlier studies devoted to the propagation of an elastic line driven in a random medium have revealed a size dependent crossover between a {\it weak pinning} regime relevant for small systems and a {\it strong pinning}  which describes the thermodynamic limit~\cite{Tanguy1998,Roux2003,Tanguy2004,Patinet2013}. For weak pinning, the macroscopic toughness is essentially the spatial average of the local toughness with a small, size dependent, correction. For strong pinning, the failure process is highly intermittent~\cite{Maloy3,Bonamy5}, and the avalanche-like dynamics results into a selection of some particularly strong regions within the fracture plane. The macroscopic toughness is thus 
larger than its spatial average.

In this work, we focus on the second case, and compute the toughness in infinitely large specimens. Our study reveals two regimes, depending on the disorder amplitude: strongly disordered materials are in the {\it individual pinning} regime, and their behavior depends on many microscopic parameters; the macroscopic toughness is shown to be dominated by the toughest obstacles. A regime of  {\it collective pinning} occurs for a lower level of disorder, where the effective toughness is shown to depend on a few measurable microscopic parameters: the average and standard deviation of the local toughness distribution and the correlation lengths of the heterogeneous toughness field.

Our theoretical description of the crack evolution relies on two competing mechanisms: the material elasticity that tends to keep the front flat and the impurities that deform it. These two mechanisms are reflected on the evolution equation of the front $u(r,t)$ that follows
\begin{equation}\label{eq_propa}
\frac{\partial u}{\partial t}(r,t)=\fext+\frac{c}{\pi}\int \frac{u(r',t)-u(r,t)}{(r'-r)^2}\dd r'-\floc(r,u).
\end{equation}
Here, the field $\floc(r,u)$ is the fluctuating local toughness
while the second term on the right hand side describes the distribution of stress intensity factor along the front resulting from its perturbed geometry~\cite{Rice4}. This integral term represents long-range elastic interactions along the fracture line and also describes the behavior of various interfaces like wetting fronts~\cite{Joanny}. The elastic constant is equal to $c= \langle \floc \rangle/2$ for fracture fronts, but we keep it as a free parameter to make our study applicable to a broader class of phenomena. Finally, $\fext$ is the macroscopic stress intensity factor prescribed by the loading conditions of the specimen, and will be also referred to as \textit{driving force} in the remainder of the article.

In Eq.(\ref{eq_propa}), the interplay between disorder and line elasticity determines the response of the line to an external force $\fext$: as long as the external force $\fext$ remains below a threshold $\fc$, the line remains \textit{pinned} by the heterogeneities; when it exceeds $\fc$, the line \textit{unpins} and acquires a non-zero asymptotic velocity~\cite{Narayan1993}. To describe this behavior, it has been particularly fruitful to consider the depinning transition as a regular critical phenomenon, with the velocity playing the role of an order parameter~\cite{Narayan1993}.  
This analogy  suggests that close to $\fc$,  the front displays an universal behavior with critical exponents and scaling laws that have been extensively investigated~\cite{Nattermann3,Narayan1993,LeDoussal4, Moulinet2004, Bustingorry}. If the test of these predictions in experiments has been rather successful~\cite{Bonamy5,Ponson14,Santucci7}, the most relevant quantity from an applied science perspective is the depinning threshold value $\fc$ which identifies with the macroscopic toughness.  Analogously to the critical temperature in equilibrium phase transition, $\fc$ is not universal and depends, to some extent, on the microscopic details of the system. The key point is how many and what features at the microscopic scale contribute in determining the value of the critical driving force. To get more insight into this transition, we also study the front roughness that is characterized using its height-height correlation function
\begin{equation}\label{def_rugo}
B(\delta r)=\overline{ \langle [u(r)-u(r+\delta r)]^2 \rangle },
\end{equation}
where $\langle \cdot \rangle$ and $\overline{\,\cdot\,}$ denote averages over the position $r$ and the disorder realization, respectively.

We describe the material microstructure by rectangular domains of constant toughness. Their length is $\xi_r$ in the $r$ direction; in the $u$ direction, it is drawn in an exponential distribution of average $\xi_u$. These rules  set a particular disorder geometry; the effect of different spatial correlations along $u$ is not  studied here and will be discussed elsewhere.
On each rectangular domain, the local toughness is $\floc=\langle \floc \rangle+\dKcloc \Floc$, where $\Floc$ is drawn from a symmetric probability distribution $P(\Floc)$ of unit variance and zero mean value.
The disorder is thus described by the standard deviation $\dKcloc$ of the toughness distribution, the correlation lengths $\xi_r$ and $\xi_u$ and the probability distribution $P(\Floc)$. The average toughness $\langle \floc \rangle$ can be absorbed in the driving force and does not play any role in the study of Eq.~(\ref{eq_propa}).
In real materials, these parameters may be estimated. For instance, in a two-phase material with toughnesses $K\ind{c}^{\mathrm{1}}$ and $K\ind{c}^{\mathrm{2}}$ and densities $n_{\mathrm{1}}$ and $n_{\mathrm{2}} = 1 - n_{\mathrm{1}}$, the disorder amplitude is $\dKcloc=\sqrt{n_{\mathrm{1}} n_{\mathrm{2}}} |K\ind{c}^{\mathrm{1}}-K\ind{c}^{\mathrm{2}}|$. The lengths $\xi_r$ and $\xi_u$ are given by the typical size of phase domains, as shown in Fig.~\ref{schema_gen}.

To identify the relevant parameters, we introduce the rescaled variables $\tilde r=r/\xi_r$, $\tilde u=u/\xi_u$ and $k\ind{ext}=(K\ind{ext}-\langle K\ind{c} \rangle)/\dKcloc$. With these variables, Eq.~(\ref{eq_propa}) reads, for a stationary state where $\partial u/\partial t=0$,
\begin{equation}\label{eq_stat_adim}
0=k\ind{ext}-\Floc(\tilde r,\tilde u(\tilde r))+\frac{1}{\pi}\frac{c\xi_u}{\dKcloc \xi_r}\int \frac{\tilde u(\tilde r')-\tilde u(\tilde r)}{(\tilde r'-\tilde r)^2}\dd\tilde r'.
\end{equation}
Thus, besides the disorder distribution $P(k\ind{c})$, the behavior of Eq.~(\ref{eq_stat_adim}) depends only on the dimensionless parameter
\begin{equation}
\Sigma=\frac{\dKcloc\xi_r}{c\xi_u},
\end{equation}
hereafter referred to as the {\it disorder parameter}.
The observables can then be written as functions of the disorder parameter and the disorder distribution multiplied by a numerical prefactor:
\begin{align}
B(r)  & = \left(\frac{\dKcloc\xi_r}{c} \right)^2 b \left(\frac{r}{\xi_r};\Sigma,P(\Floc) \right)  \label{eq_scaling_roughness}\\
\tfc & = \fc-\langle \floc \rangle = \dKcloc k\ind{eff}(\Sigma, P(k_\mathrm{c})).
\end{align}

To perform numerical simulations of Eq.~(\ref{eq_propa}), the system is discretized in the $r$ direction with a step $\xi_r$ and put on a strip of finite width $L$, with periodic boundary conditions. The driving force is replaced by a parabolic drive centered at $w$ and of curvature $\kappa$: $K\ind{ext}=\kappa[w-u(r,t)]$.  We start with a flat configuration $u(r)=0$ and set $w=w_0>0$. The line advances to a stable state $u_{w_0}(r)$ that can be found using the algorithm proposed in~\cite{Rosso2}. 
The stable configuration does not depend on the dynamics as soon as it satisfies the Middleton no passing rule~\cite{Middleton2}; our dynamics, detailed in~\cite{Rosso2}, satisfies this rule and is designed to converge rapidly to the first stable configuration.
This configuration depends in general on the initial configuration. However,  according to the no passing rule theorem~\cite{Middleton2}, there exists $w^*$ such that for any $w_0>w^*$,  $u_{w_0}(r)$  becomes independent of the initial configuration. The pinning force acting on a stable line is measured as 
\begin{equation}
K_{w_0}(\kappa)=\kappa\left[w_0- \left\langle u_{w_0}(r) \right\rangle\right].
\end{equation}
Moving the parabola, we have computed $1000$ stable and well separated configurations so that their respective pinning forces are uncorrelated variables of average $\overline{K(\kappa)}$ and variance $\overline{\delta K(\kappa)^2}$.

The statistical tilt symmetry assures that the linear part of the equation of motion (and thus the curvature $\kappa$) is not renormalized~\cite{Schulz1988}. This means that the length associated with the curvature $L_\kappa=c/\kappa$ fixes the distance from the critical point (located at $\fc$) as $\kappa^{1-\zeta}$. When $\kappa\to 0$ while $L\gg L_\kappa$, the average pinning force tends to the thermodynamical value,  $\fc$,  with finite size effects of the form 
$\overline{K(\kappa)}=\fc +c_1 \kappa^{1-\zeta}+\cdots$~\cite{Rosso2007}.
In the limit $L\gg L_\kappa=c/\kappa$, the interface can be modeled as a collection of independent interfaces of size $L_\kappa$ and the central limit theorem assures that the variance $\overline{\delta K(\kappa)^2}$ should  scale as $\sim (L_\kappa/L) \kappa^{2(1-\zeta)}$. This allows to write the finite size effects without the exponent $\zeta$,
\begin{equation}
\label{finitesize}
\overline{K(\kappa)}=\fc +c_1 \sqrt{\kappa L\overline{\delta K(\kappa)^2}}+\cdots.
\end{equation}
The value  $\fc$ is extrapolated using Eq.~(\ref{finitesize}) and samples of size $L=1000$ with $\kappa$ ranging from $10^{-1}$ to $10^{-3}$. 
The roughness $B(r)$ is obtained by  averaging over $100$ stable configurations with $L=10^4$ and $\kappa=10^{-3}$.
Note that other finite size procedures allow to extrapolate $\fc$. For example one can compute the value of the critical force in samples of size  $L \times L^\zeta$ and then take the limit $L\to \infty$. 
It has been shown recently that all these methods converge to the same value of $\fc$~\cite{Kolton2013}, that is an intrinsic feature of the material. Here we use four different disorder distributions $P(\Floc)$: bivalued, "rare" (with density $n=0.1$) \footnote{The rare distribution with a density $n$ is defined by $P_{\textrm{rare},n}(\Floc) = (1-n)\delta(\Floc) + \frac{n}{2}\left[\delta\left(\Floc+\frac{1}{\sqrt{n}}\right)+\delta\left(\Floc-\frac{1}{\sqrt{n}}\right)\right]$.}, Gaussian, and exponential. $\xi_u$ is varied from $0.001$ to $30$ and $\dKcloc$ from $1$ to $8$, and we set $c=1$ and $\xi_r=1$.

\begin{figure}
\includegraphics[width=\linewidth]{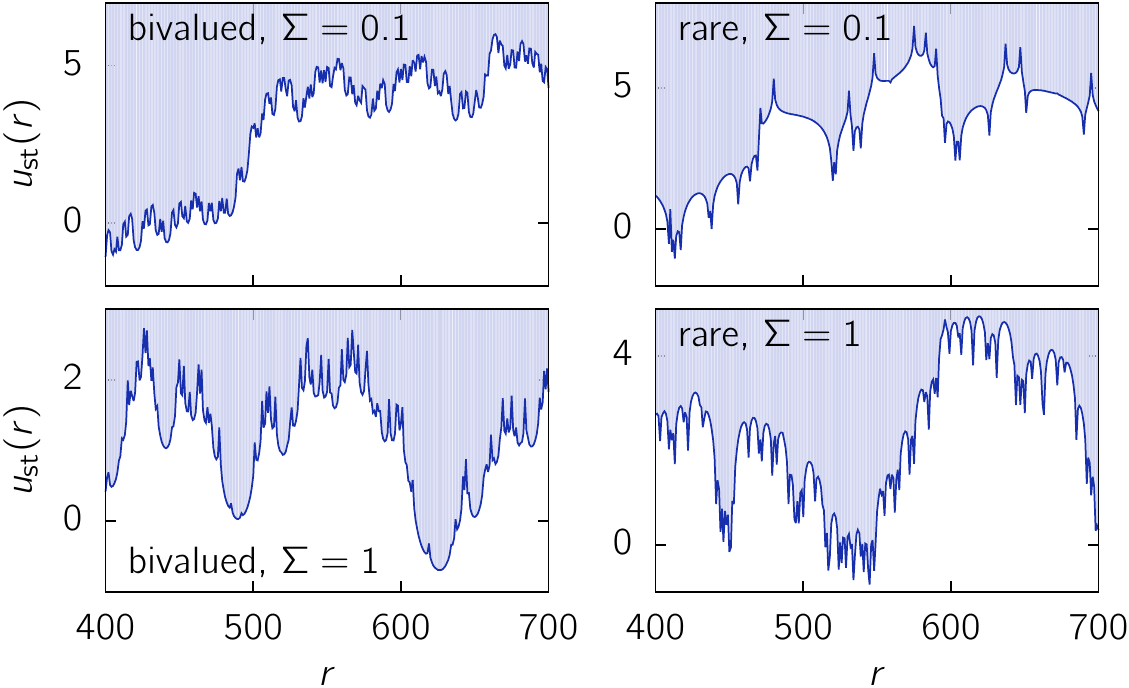}
\caption{(Color online) Stable front configurations for bivalued and rare disorder distributions and two values of the disorder parameter $\Sigma$. The fronts propagate upwards and the $u$ coordinate has been shifted.}
\label{fig_fronts}
\end{figure}

\begin{figure}
\begin{center}
\includegraphics[width=.9\linewidth]{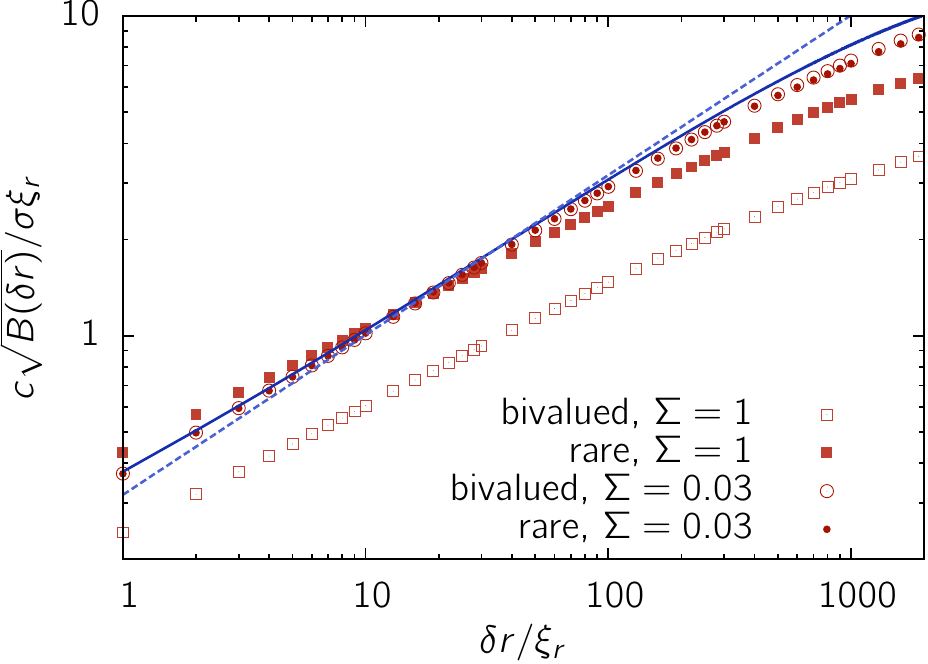}
\end{center}
\caption{(Color online) Crack front roughness: comparison between a bivalued (open symbols) and a rare distribution (full symbols), at low (circles) and high (squares) disorder. The solid line is the prediction (\ref{eq_lark_rugo}) using Larkin's approximation and the dashed line the long distance approximation~(\ref{ld_lark_rugo}).}
\label{fig_rugo}
\end{figure}

Examples of stable front configurations are shown in Fig.~\ref{fig_fronts}, for bivalued and rare disorder distributions. At high disorder ($\Sigma=1$), the configurations look similar but the front amplitude is larger for the rare distribution. 
On the contrary, at low disorder ($\Sigma=0.1$), the configurations look very different but their amplitudes are the same.
The corresponding roughness functions are given in Fig.~\ref{fig_rugo}. For a high disorder, the roughness is much higher for a rare disorder than for a bivalued disorder: this agrees with the direct observation of the fronts. For a low disorder, the roughness does not seem to depend on the disorder distribution $P(\Floc)$. The difference in the front shape is not seen in the roughness; this is not surprising since the roughness, as a two-point correlation function, cannot give a complete account of the front shape.

\begin{figure}
\begin{center}
\includegraphics[width=.9\linewidth]{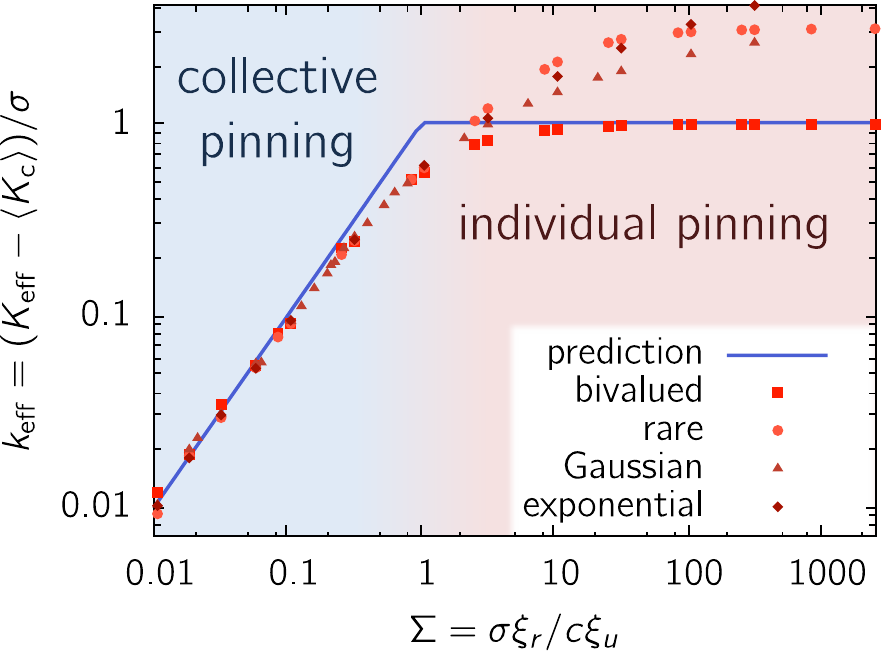}
\end{center}
\caption{(Color online) Dimensionless disorder induced toughening $\tFc$ for different disorder distributions as a function of the disorder parameter $\Sigma$. The line shows the prediction (\ref{fc_lark}).}
\label{fig_fc}
\end{figure}

The disorder induced toughening (or net critical driving force) $\tfc=\fc-\langle \floc \rangle$ is plotted versus the disorder parameter $\Sigma$ on Fig.~\ref{fig_fc}; two regimes can be distinguished. At low disorder ($\Sigma<1$), we observe a beautiful collapse between different disorder distributions: this is the {\it collective pinning} regime. The net critical driving force follows the phenomenological law
\begin{equation}\label{fc_pheno}
\tfc\simeq\frac{\dKcloc^2\xi_r}{c\xi_u}=\dKcloc\Sigma \, .
\end{equation}
This is our main result, justified later with physical arguments.

On the contrary, at high disorder ($\Sigma>1$), the net critical driving force depends strongly on the disorder distribution: this is the {\it individual pinning} regime.
In the limit $\Sigma\rightarrow\infty$, that can be interpreted as the limit $c\rightarrow 0$, the front is softer and has access to more and more disorder realizations. It may thus "choose" the most pinning one. As a result, when the disorder is a bounded distribution, the net critical driving force gets close to the bound. It is indeed what is observed for the bivalued distribution, whose maximum is $1$, and for the "rare" distribution, whose maximum is $1/\sqrt{n}\simeq 3.16$. If the disorder is not bounded, the critical driving force is likely to diverge, as observed in the simulations. The fatter the distribution tail is, the faster the divergence is expected, that is also what is observed: the divergence is faster for the exponential distribution than for the Gaussian one.

We now show that physical arguments adapted from the Larkin and Ovchinnikov study of vortex pinning in superconductors~\cite{Larkin1970,Larkin1979} allow for an intuitive interpretation of our results. The main difficulty in dealing with Eq.~(\ref{eq_propa}) is its non linearity coming from the disorder term. The first Larkin assumption is that this difficulty can be circumvented at {\it short distances}, where the crack front does not see that the disorder correlation length $\xi_u$  is finite. 
The Larkin model is thus defined as the limit $\xi_u=\infty$ ({\it i.e.} $\Sigma=0$), that amounts of removing the $u$ dependence of the disorder: $\floc(r,u)\rightarrow \floc(r)$~\cite{Larkin1970}. 

To allow comparison with the numerical simulations, we write the Larkin model for a line of length $L$ discretized with a step $a$, with periodic boundary conditions and put in a parabola of curvature $\kappa$ centered at $w=0$.
A stationnary solution satisfies
\begin{equation}\label{eq_larkin_discret}
0=-\kappa u_j+\frac{c}{\pi a}\sum_{j'\neq j}\phi_{j-j'}(u_{j'}-u_j) - \floc_j,
\end{equation}
where
\begin{equation}
\phi_j=\sum_n\frac{1}{(j-nL/a)^2}
\end{equation}
is the elastic kernel taking into account the periodic boundary conditions. Now, we Fourier transform this equation using 
$\tilde u_k = \sum_{j=0}^{N-1} \exp\left(-2\pi i \frac{jk}{N}\right) u_j$, 
where $N=L/a$ is the number of points of the line; the stationary condition (\ref{eq_larkin_discret}) now reads
\begin{equation}
-\kappa \tilde u_k+\frac{c}{\pi a}\left(\tilde\phi_k-\tilde\phi_0\right)\tilde u_k=-\tilde\floc_k.
\end{equation}
Extracting $\tilde u_k$ and computing the average $\overline{\tilde u_k\tilde u_{k'}}$ gives the roughness as a function of the disorder correlation.
We choose a disorder correlation $\overline{\floc_j\floc_{j'}}=\sigma^2\delta_{j,j'}$; in this case the discretization length corresponds to the correlation length, $a=\xi_r$.
The roughness at a distance $\delta r=a j$ is then
\begin{equation}\label{eq_lark_rugo}
B\ind{Larkin}(a j)=\left(\frac{\sigma a}{c} \right)^2  \frac{2\pi^2}{N}\sum_k \frac{1-\cos \left(\frac{2\pi j k}{N} \right)}{\left(\tilde\phi_0-\tilde\phi_k +\frac{\pi\kappa a}{c} \right)^2};
\end{equation}
in the limit $\kappa\to 0$, $N\to\infty$, it becomes
\begin{equation}\label{eq_lark_rugo_cont}
B\ind{Larkin}(\delta r)=\left(\frac{\sigma}{c} \right)^2  \frac{a}{\pi}\int_{-\pi/a}^{\pi/a} \frac{1-\cos(k\delta r)}{k^2\left(1-\frac{a|k|}{2\pi}\right)^2}\dd k.
\end{equation}
For distances short enough for the Larkin model to be relevant, but large compared to the disorder correlation length, $\delta r\gg a=\xi_r$, the roughness reduces to
\begin{equation}\label{ld_lark_rugo}
B\ind{Larkin}(\delta r\gg\xi_r)\sim \left(\frac{\dKcloc\xi_r}{c} \right)^2 \frac{\delta r}{\xi_r},
\end{equation}
giving the Larkin roughness exponent $\zeta\ind{Larkin}=0.5$.

To explore the roughness below the correlation length $\xi_r$, we need to take a disorder correlation function with correlation length larger than the discretization step, $\xi_r\gg a$.
In practice, this cuts the integral in (\ref{eq_lark_rugo_cont}) at $\pi/\xi_r$, with an extra factor $\xi_r/a$. In the limit $\delta r\ll\xi_r$, it reduces to
\begin{equation}\label{eq_lark_rugo_bal}
B\ind{Larkin}(\delta r\ll\xi_r)\sim \left(\frac{\sigma}{c} \right)^2\delta r^2:
\end{equation}
the line has a ballistic behavior at short distances.

The exact Larkin roughness (\ref{eq_lark_rugo}) and its limiting law (\ref{ld_lark_rugo}) are compared in Fig.~\ref{fig_rugo} with the numerical simulations for finite $\xi_u$. For low disorder, the agreement between simulations and the Larkin prediction is strikingly good. The long distance approximation is also rather good. On the contrary, as expected,  high disorder results are far from the prediction.

We can address the question of the validity of the Larkin regime: what does the assumption of "short distances" mean? The answer involves the roughness that depicts the amplitude of the front perturbations at different scales. If these perturbations are smaller than the correlation length $\xi_u$, the Larkin model should represent correctly the behaviour of the line; on the contrary, when they are larger than $\xi_u$, a different behaviour is expected. This defines the length in the $r$ direction up to which the Larkin approximation is relevant, the so-called {\it Larkin length} $\Lc$ (see Ref.~\cite{Agoritsas} for a recent review), by
\begin{equation}
\sqrt{B\ind{Larkin}(\Lc)}=\xi_u.
\end{equation}
If the Larkin length is larger than the correlation length $\xi_r$, its expression can be derived from Eq.~(\ref{ld_lark_rugo}):
\begin{equation}\label{eq_larkin_length}
\Lc=\left(\frac{c\xi_u}{\dKcloc\xi_r} \right)^2\xi_r=\frac{\xi_r}{\Sigma^2}.
\end{equation}
The domains of size $\Lc$ that behave as if $\xi_u$ was infinite are called {\it Larkin domains}; in these domains and for distances larger than $\xi_r$, the roughness is given by (\ref{ld_lark_rugo}). 
For larger distances, the roughness departs from the Larkin prediction and continuity implies
\begin{equation}\label{eq_rugo_zeta}
B(\delta r\gg\Lc)\simeq \xi_u (\delta r/\Lc)^\zeta,
\end{equation}
with a roughness exponent $\zeta\simeq 0.39$~\cite{Rosso2}.

The second Larkin assumption is that the critical driving force is given by the typical toughness seen by a Larkin domain. The toughness averaged along the line on a Larkin domain has a mean $\langle \floc \rangle$ and its standard deviation depends on $\Lc$: if $\Lc<\xi_r$, the domain sees only one defect and the standard deviation is $\dKcloc$; if $\Lc>\xi_r$, the local toughness is averaged over $\Lc/\xi_r$ uncorrelated defects and the standard deviation is $\dKcloc\sqrt{\xi_r/L_r}$. As a consequence:
\begin{itemize}
\item if $\Lc>\xi_r$ (or $\Sigma<1$), a Larkin domain sees several defects: this is the {\it collective pinning} regime, where the net critical driving force is $\tfc=\dKcloc\sqrt{\xi_r/\Lc}$.
\item if $\Lc<\xi_r$ (or $\Sigma>1$), a Larkin domain sees only one defect: this is the {\it individual pinning} regime. The net critical driving force is now $\tfc=\dKcloc$.
\end{itemize}
Using the long distance expression of the Larkin length (\ref{eq_larkin_length}), we can write explicitly the net critical driving force,
\begin{equation}\label{fc_lark}
\tfc=\left\{ \begin{array}{cllll}
\displaystyle{\frac{\dKcloc^2\xi_r}{c\xi_u}} & = & \dKcloc\Sigma & \textrm{if } \displaystyle{\Sigma=\frac{\dKcloc\xi_r}{c\xi_u}<1},\vspace{2mm}\\
\dKcloc &  &  & \textrm{if } \displaystyle{\Sigma=\frac{\dKcloc\xi_r}{c\xi_u}>1}.
\end{array}\right.
\end{equation}

These expressions are compared to the simulations in Fig.~\ref{fig_fc} for the four different disorder distributions. In the collective pinning regime, the prevision captures the results of the simulations: this confirms the validity of Eq.~(\ref{fc_pheno}), irrespective of the underlying local toughness distribution. The effective toughness, that is a priori a non-universal quantity, is shown here to be remarkably robust.
In the individual pinning regime, the prediction (\ref{fc_lark}) does not fully describe the results of the simulations, but provides a lower-bound of the effective toughness
\begin{equation}
\fc\geq\langle \floc \rangle + \dKcloc.
\end{equation}
Indeed, in this regime, the toughest defects which set the value of the critical force are systematically tougher than the toughness standard deviation $\sigma$.

As a possible application, we consider the case of a material of toughness $K\ind{c}^1$ reinforced by a small volume fraction $n_2\ll 1$ of randomly distributed particles with large toughness $K\ind{c}^2$ and define the contrast as $C=(K\ind{c}^2-K\ind{c}^1)/K\ind{c}^1$. Considering an isotropic distribution of particles ($\xi_r=\xi_u$) for simplicity, Eq.~(\ref{fc_lark}) predicts an increase of the effective toughness of the material by a factor $\fc/K\ind{c}^1 \simeq 1+n_2 (C+2 C^2)$ in the collective regime. This result is quite different from the predictions of Refs.~\cite{Gao,Bower} obtained for a periodic array of tough particles for which $\fc / K\ind{c}^{1} \simeq 1 + 2 C \sqrt{n_2/\pi}$, highlighting the crucial role played by disorder.

We underline the limitations of the line model used here in the context of fracture. The line model rests on two assumptions: (i) The solid is brittle, {\it i.e.}~the process zone is small compared to the microstructural length scales $\xi_r$ and $\xi_u$. This allows to treat the crack front as a line that separates the fracture plane in two distinct domains of intact and broken material.
(ii) The distortion of the line is small enough so that the linearized interaction term used in Eq.~(\ref{eq_propa}) describes properly the line elasticity. 
This condition is fullfilled as long as the steepest slopes along the crack line are smaller than one. 
The steepest slopes are found at small distances, for $\delta r \ll \xi_r$: the ballistic roughness (\ref{eq_lark_rugo_bal}) shows that they are $[\sqrt{B(\delta r)}/\delta r]\ind{max}\simeq \sigma/c$.
The condition $\sigma/c\leq 1$, or $\sigma\leq \langle K\ind{c} \rangle/2$, limits our study to moderately heterogeneous solids, where the spatial toughness variations are smaller than the average toughness. This condition is generally satisfied in the collective pinning regime whereas in the individual pinning regime where $\sigma/c>\xi_u/\xi_r$ (Eq.~\ref{fc_lark}), it can be only satisfied for anisotropic disorder with $\xi_r>\xi_u$. Beyond the assumption (ii), larger local slopes of the order of one would require non-linear corrections to the line elasticity~\cite{Vasoya}. However, we expect a similar toughening mechanism governed by the toughest defects to apply. If the steepest slopes are even larger, the line model (assumption (i)) breaks down: this may come from crack bridging due to very tough inclusions~\cite{Budiansky,Bower} that conserve clusters of intact materials in the broken domain; or from the nucleation and coalescence of damage ahead of the crack that occurs {\it e.g.}~in quasi-brittle solids. In this latter case, the front becomes fractal~\cite{Gjerden} and a material description based on random networks of fuses or springs will be more appropriate~\cite{Moreira2012}. 

We now discuss how the regimes identified here compare with those identified in former studies.
In the self coherent schemes~\cite{Roux2003, Patinet2013}, the weak to strong pinning transition corresponds to the emergence of metastability: for small systems or low disorder, there is no metastability and the effective toughness is equal to the space averaged toughness; for large systems or strong disorder, metastability produces a disorder induced toughening. The transition depends on the system size and the weak pinning regime does not survive in the thermodynamic limit. In our case, metastability is always present, indicating that the individual and collective pinning regimes are part of the strong pinning regime. The individual to collective pinning crossover occurs when the Larkin length $L\ind{c}$ equals the disorder correlation length $\xi_r$. 
It is then natural to expect that the weak to strong pinning transition occurs when the Larkin length equals the system size $L$. In the weak pinning regime $L\ind{c}>L$, the system size sets the size of the Larkin domains, from which it follows
\begin{equation}\label{eq_larkin_weak}
\tfc=\sigma \sqrt{\frac{\xi_r}{L}}.
\end{equation}
This regime occurs for very low disorder $\sigma < c \xi_u/\sqrt{L \xi_r}$ and has a small effect on the material toughness; this explains why it has been neglected in some studies~\cite{Roux2003, Patinet2013}.

In the linear stability analysis~\cite{Tanguy2004}, the different behaviors show up in the \textit{participation ratio} $\tau$ of the most unstable mode.
In the weak pinning regime, the system unpins as a whole and $\tau\simeq 1$; in the strong pinning regime, the avalanches are concentrated on a small portion of the system and $\tau < 1$.  Following Larkin's argument, the Larkin length would also be the minimal size of avalanches and would thus be linked to the participation ratio via $\tau=L\ind{c}/L$. With this interpretation, a participation ratio in the range $ \xi_r/L < \tau < 1 $ would reveal a collective regime, while $\tau\simeq \xi_r/L$ would be the signature of individual pinning. 
However, the connection between the correlation length of the most unstable mode and the Larkin length deserves further investigations, as emphasized in~\cite{Tanguy2004}. 

Finally, our study provides also insights on the behavior of contact lines that follow the same evolution equation as crack lines. In Refs.~\cite{DiMeglio1992, Crassous1994}, the disorder amplitude can be computed from the defects density $n$ and strength $h$ as $\sigma^2\sim nh^2$ while the capillary length sets the system size $L$. The hysteresis $H$ in wetting problems is analogous to the disordered induced toughening $\tfc$ for cracks. First, it was found that the hysteresis is zero for a defect strength $h$ below a critical value $h\ind{c}$; that corresponds to the weak pinning regime. As noted in~\cite{Crassous1994}, this regime disappears when the capillary length goes to infinity. In addition, the scaling $h\ind{c}(n) \sim 1/\sqrt{n}$ is consistent with the weak to strong pinning crossover occurring for $L = L_\mathrm{c} = c^2\xi_r/(n h^2)$, as suggested here. Above this critical value, but at low densities, the hysteresis scales as $H\sim n h^2\sim\sigma^2$: this behavior is consistent with the collective pinning regime in Eq.~(\ref{fc_lark}).

To conclude, we have investigated the effect of a disordered microstructure on the pinning of a brittle crack front. We have shown that the interaction between the front and the disorder is governed by one dimensionless parameter and that two distinct regimes can be identified. In the collective pinning regime that occurs at low disorder amplitude, many impurities act together to pin the crack front. The effective toughness follows Eq.~(\ref{fc_pheno}) and depends on a few parameters measurable from the material microstructure. On the other hand, at high disorder amplitude, the front is pinned by tough individual defects: this is the individual pinning regime. The macroscopic toughness is then shown to depend on many parameters, such as the actual distribution of microscopic local toughness. 

\acknowledgments
The authors would like to thank J.-B. Leblond, V. Lecomte, N. Pindra and S. Roux for fruitful discussions.

\end{document}